\documentclass[twocolumn,prb,amsmath,amssymb,showpacs,superscriptaddress,floatfix]{revtex4}
\usepackage{graphicx}
\usepackage{epsfig}
\usepackage{amsbsy}

\makeindex

\begin{document}

\title{Andreev bound states and current-phase relations in three-dimensional topological insulators}
\author{M. Snelder, M. Veldhorst, A. A. Golubov, A. Brinkman}
\affiliation{Faculty of Science and Technology and MESA+ Institute for Nanotechnology, University of Twente, 7500 AE Enschede, The Netherlands}
\date{\today}

\begin{abstract}
To guide the search for the Majorana fermion, we theoretically study  superconductor/topological/superconductor (S/TI/S) junctions  in an experimentally relevant regime. We find that the striking features present in these systems, including the doubled periodicity of the Andreev bound states (ABSs) due to tunneling via Majorana states, can still be present at high electron densities. We show that via the inclusion of magnetic layers, this 4$\pi$ periodic ABS can still be observed in three-dimensional topological insulators, where finite angle incidence usually results in the opening of a gap at zero energy and hence results in a 2$\pi$ periodic ABS. Furthermore, we study the Josephson junction characteristics and find that the gap size can be controlled and decreased by tuning the magnetization direction and amplitude. These findings pave the way for designing experiments on S/3DTI/S junctions. 

\end{abstract}
\maketitle
The prediction of Fu and Kane \cite{Fu2008} that Majorana fermions can be realized in superconductor$-$3D topological insulator structures, boosted theoretical predictions for the peculiar Majorana fermion properties \cite{ Nilsson2008, Tanaka2009, Hasan2010, Alicea2012,beenakkerreview}. Strong progress has been made in the fabrication of two-dimensional \cite{Koenig, Koenig2, Roth, Tkachov} and three-dimensional topological insulators \cite{Tkachov, Zhang, Mele, Fu2007, Murakami, Moore, Hsieh2008, Hsieh20092, Tanaka2012}. Recently, the three-dimensional topological insulators (TIs) based on the Bi-compounds (e.g. Bi$_{2}$Te$_{3}$, Bi$_{2}$Se$_{3}$)  have already led to the realization of superconductor(S)/TI/S junctions \cite{Samarth, Williams, Morpurgo, Veldhorst,Wang2012,Cho2012} and SQUIDs \cite{Veldhorst2, Qu}. From an experimental point of view it is difficult to realize topological insulator materials with the chemical potential at or close to the Dirac point. It is therefore highly desirable to have a guiding theory in an experimentally relevant regime that can pave the way towards the verification of the Majorana fermion in S/TI hybrids. 

\begin{figure}[t]
	\centering 
		\includegraphics[width=0.5\textwidth]{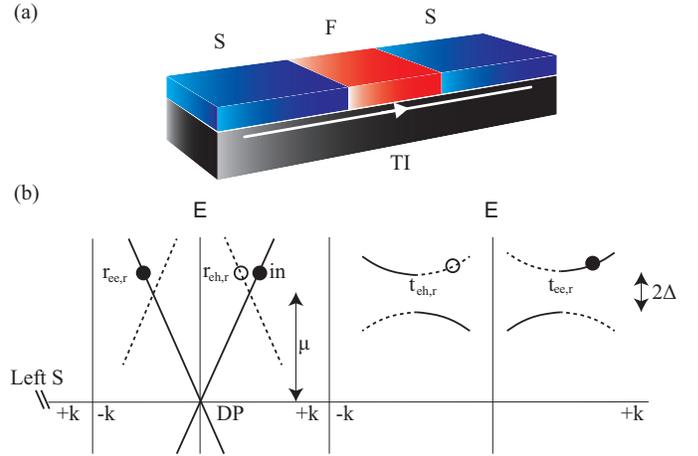}
		\caption{(a) Schematic drawing of a topological insulator (TI) with a superconductor (S) and ferromagnet (F) on top. We consider the bound states at the surface of the TI with the proximity effect from both the superconductor and ferromagnet. (b) The energy dispersion at the TI surface (left) and at the S side (right). DP indicates the Dirac point. Here the case is shown without a F and with an incoming electron at the right interface.}  
		\label{fig:1}
		\vspace{-15pt} 
\end{figure}
\begin{figure*}[t!]
\centering
		\includegraphics[width=0.75\textwidth]{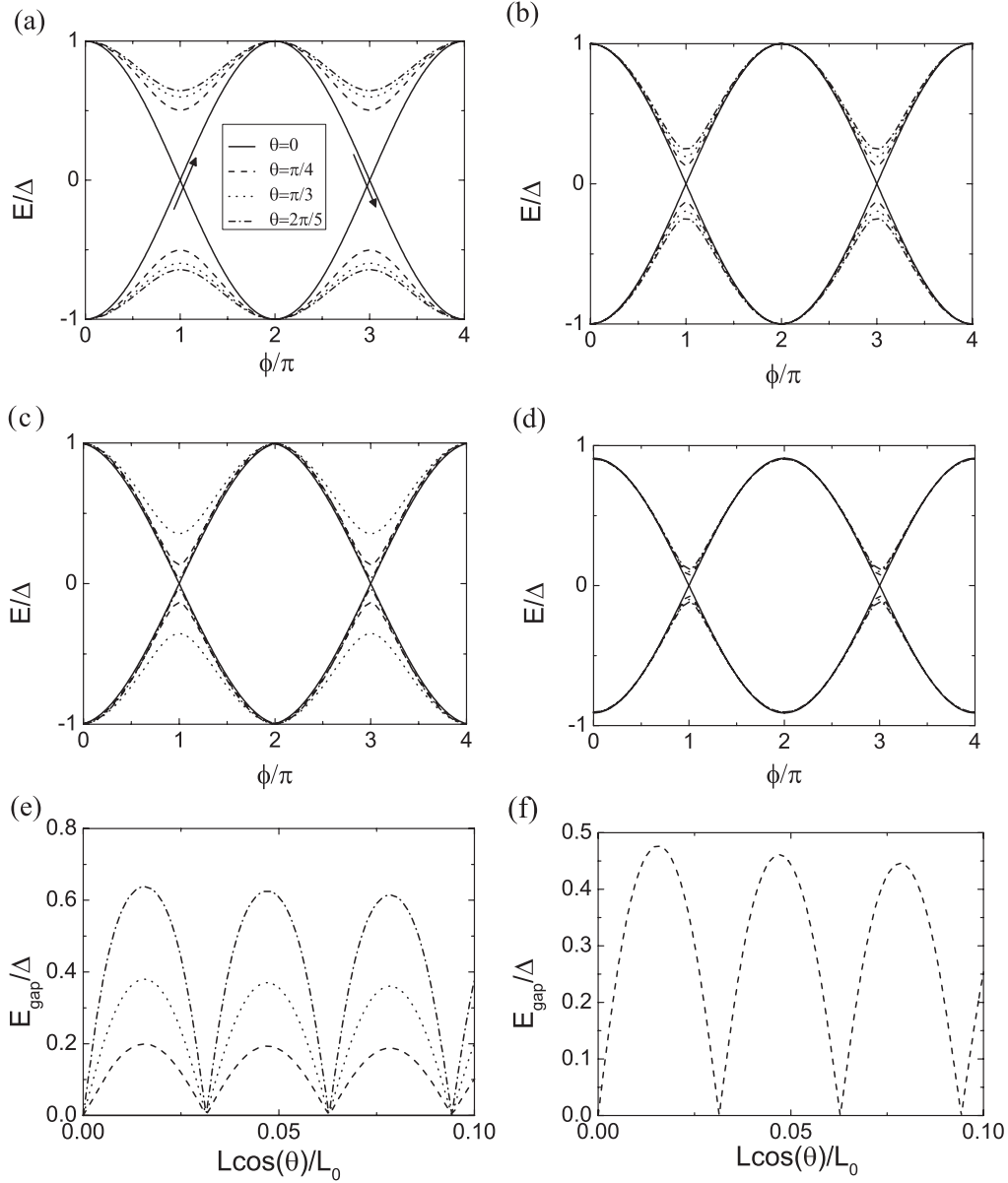}
		\caption{(a) Andreev bound states for different trajectories in a S/TI/S junction. A gap opens at finite angle. The arrows indicate which bound state branches are connected to form a 4$\pi$ periodicity. In these branches a Majorana fermion is present at $\phi=\pi$. The legend applies to all the figures. $\mu_{TI}/\Delta=100, \mu_{S}/\Delta=1000$ and $L/L_{0}=0.01$ where $L_{0}=v_{f}\hbar / \mu_{TI}$. (b) $\mu_{TI}/\Delta=100, \mu_{S}/\Delta=120$ and $L/L_{0}=0.01$. (c) $\mu_{TI}/\Delta=100, \mu_{S}/\Delta=120$ and $L/L_{0}=0.1$ and (d) $\mu_{TI}/\Delta=100, \mu_{S}/\Delta=120, m_{z}/\Delta=60.0$ and $L/L_{0}=0.01$. (e) Bound state energy for different angles and phase difference $\phi=\pi$. Furthermore $\mu_{S}/\Delta=120, \mu_{TI}/\Delta=100, m_{z}/\Delta=0$. The energy is oscillating with length. E$_{gap}$ is defined as the distance from E/$\Delta=0$ till the minimum of the ABS. (f) Bound state energy for fixed phase difference $\phi=\pi$ and angle using the formula from Ref. \onlinecite{Furusaki1991}.}
		\label{fig:2}
		\vspace{-15pt} 
\end{figure*}

Here we theoretically study superconductor/three-dimensional topological insulator Josephson junctions. In the calculations of Ref. \onlinecite{Tanaka2009, Fu2008, Akhmerov2009, Fu2009, Linder2010} it is assumed that the Fermi level is close to the Dirac point. In addition, it is always assumed that the ferromagnet placed on top of a TI, has a magnetization $|M|>\mu$ in these calculations. In these systems, an Andreev bound state (ABS) with a doubled periodicity is predicted. We consider the experimentally relevant regime of high electron densities, and show that, despite the chemical potential $\mu$ being situated far away from the Dirac point, this characteristic feature is still present. We furthermore consider the presence of a ferromagnetic layer with magnetization $|M|<\mu$ on top of the junction, and show that it can drastically alter the Josephson characteristics, even when $(|M|,\Delta) \ll \mu$. This is particularly interesting, since the magnetization opens a gap not at the Fermi energy, as the superconducting correlations do, but at the Dirac point, far away from $E_{F}$. We show that a gap in the superconducting bound states spectra always opens at a finite angle of incidence. However, the size of this gap can be tuned and decreased, and can in principle vanish, by increasing the perpendicular magnetization amplitude. 

The discussion of the bound states of a S/TI/S junction in this article is organized as follows: first, we study the case without a ferromagnetic layer on top of the TI. Then we discuss the bound states with a ferromagnet. We will see that a 4$\pi$ periodic Andreev bound state is still present in the 3D case but only for one channel. This 4$\pi$ periodic ABS is a feature of the presence of Majorana fermions. We show the supercurrent obtained by the bound states and discuss the observation of a 4$\pi$ periodic ABS in a 3D topological insulator.

\section{The S/3DTI/S junction}
The configuration of the junction we consider, is as shown in Fig. \ref{fig:1}(a). In the Nambu basis
\begin{eqnarray}
\Psi=\left( \psi_{\upharpoonleft},\psi_{\downharpoonleft},\psi_{\upharpoonleft}^{\dagger},\psi_{\downharpoonleft}^{\dagger}\right) ^{T}
\end{eqnarray}
the Hamiltonian with a superconducting and magnetic proximity effect is\cite{Linder2010}
\begin{eqnarray}
H&=&\left(\begin{array}{cc}
H_{0}(\textbf{k})+M & \Delta(\textbf{k})\\
-\Delta^{*}(-\textbf{k}) & -H_{0}^{*}(-\textbf{k})-M^{*} 
\end{array} \right), 
\end{eqnarray}
where 
\begin{eqnarray}
H_{0}(\textbf{k})=v_{F}(\sigma_{x}k_{x}+\sigma_{y}k_{y})-\mu_{j}, \hspace{6mm} M=\textbf{m}\cdot \boldsymbol\sigma,
\end{eqnarray}
with $M$ the magnetization due to the ferromagnet and $\textbf{m}=\left( m_{x}, m_{y}, m_{z} \right)$ the exchange field. $\sigma_{i}$ are the Pauli matrices and the index $j$ of the chemical potential is S for in the superconducting part and TI in the topological insulator. From this Hamiltonian we calculated the eigenvectors of the TI in the presence of the proximimty effects. We assume the superconductor to be an s-wave superconductor, $\Delta(\textbf{k})=\Delta e^{i\phi_{S}}$, where $\phi_{s}$ is the superconducting phase and $\Delta=0$ outside the superconductor. This Heaviside function of the order parameter simplifies the calculations. However, it is not expected that there will be a qualitatively difference with the results when a self-consistent order parameter is used. As for example is shown in Ref. \onlinecite{Barash1997} for a d-wave superconductor where the induced order parameter is taken into account in the normal metal, the Andreev bound state spectrum does not change near zero-energy. It is this regime in the spectrum that is relevant for the appearance of a Majorana mode. 

The non-superconducting part can be under the influence of a ferromagnetic proximity effect. In Majorana devices the magnetization is taken perpendicular to the TI surface, $M= m_{z} \sigma_{z}$ (the resulting eigenvectors are listed in the appendix). Magnetization parallel to the interface causes only a shift in the wave vector but does not open a gap. In the known topological insulators, the Dirac point is in the middle of the band gap or close to the valence band. The chemical potential is usually close to the conduction band as the topological insulators based on Bi-compounds are not really good insulators yet. Therefore, the chemical potential, $\mu_{TI,S}$, is much larger than the superconducting gap in experiments. In this case we have only normal Andreev reflection at the interface and no specular Andreev reflection \cite{BeenakkerGraphene}. 

For most proposals it is desired to have a ferromagnet on top of the TI (preferably an insulator so that practically no current flows through the ferromagnet)\cite{Tanaka2009, Akhmerov2009, Fu2009, Linder2010}. We estimate here how much the gap at the Dirac point can be opened by such a ferromagnet. For a magnetic moment of $n\mu_{B}$ per unit cell of size $a^{3}$, where $n$ is an integer and $\mu_{B}$ is the Bohr magneton, we can estimate the value of the $m_{z} \sigma_{z}$ part of the Hamiltonian. Making the assumption that the atoms can be approximated by spheres with $n$ elementary dipoles  and perfect coupling to the TI, we estimate that the opened gap will be about 0.0002$n/a^{3}$ eV where $a$ is in \AA ngstr\"{o}m. This value is typically smaller than the value of the Fermi energy inside the gap ($>0.05$ eV). We therefore study the relevant regime of $m_{z}<\mu_{TI,S}$.

With these assumptions we solve the Andreev and normal reflection coefficients at both left and right interfaces for incoming electrons and holes by matching the wave functions at the interface. Following Kulik\cite{Kulik} (see example in Ref. \onlinecite{Zagoskin}), the wave function in the topological insulator can be written as
\begin{eqnarray}
\psi=a\psi_{e}^{+}+b\psi_{h}^{+}+c\psi_{e}^{-}+d\psi_{h}^{-},
\end{eqnarray} where the index $e$ and $h$ indicate an electron and hole wave function respectively. The coefficients $a, b, c, d$ are related by
\begin{eqnarray}\label{eq1:zagoskin}
c\psi_{e}^{-}e^{-i|\textbf{k}_{e}|\tilde{L}}&=& 
r_{ee, r}a\psi_{e}^{+}e^{i|\textbf{k}_{e}|\tilde{L}}+r_{he, r}d\psi_{h}^{-}e^{-i|\textbf{k}_{h}|\tilde{L}}, \nonumber \\
b\psi_{h}^{+}e^{i|\textbf{k}_{h}|\tilde{L}}&=&
r_{eh, r}a\psi_{e}^{+}e^{i|\textbf{k}_{e}|\tilde{L}}+r_{hh, r}d\psi_{h}^{-}e^{-i|\textbf{k}_{h}|\tilde{L}}, \nonumber \\
a\psi_{e}^{+}e^{-i|\textbf{k}_{e}|\tilde{L}}&=&
r_{ee, l}c\psi_{e}^{-}e^{i|\textbf{k}_{e}|\tilde{L}}+r_{he, l}b\psi_{h}^{+}e^{-i|\textbf{k}_{h}|\tilde{L}},\nonumber \\
d\psi_{h}^{-}e^{i|\textbf{k}_{h}|\tilde{L}}&=&
r_{eh, l}c\psi_{e}^{-}e^{i|\textbf{k}_{e}|\tilde{L}}+r_{hh, l}b\psi_{h}^{+}e^{-i|\textbf{k}_{h}|\tilde{L}},
\end{eqnarray} where $\tilde{L}=\frac{L}{2}\cos \theta$. The second indices of the wave functions refer to the right (r) and left (l) interface. The right interface is placed at $L/2$ and the left interface at $-L/2$. $\theta$ is the angle of incidence from the TI to the S where zero angle means orthogonal to the interface. The wave vectors $k_{h}$ and $k_{e}$ are the wave vectors of the hole and electron respectively. They are given by
\begin{eqnarray} 
|\textbf{k}_{h}|&=&\sqrt{\left(\mu_{TI}-E \right)^{2} -m_{z}^2}/v_{F},\nonumber \\
|\textbf{k}_{e}|&=&\sqrt{\left(\mu_{TI}+E \right)^{2} -m_{z}^2}/v_{F},
\end{eqnarray} 
where $v_{F}$ is the Fermi velocity. The mismatch between these wave vectors and the wave vector in the superconductor causes an effective barrier at the interface. The superconducting gap can be neglected in this mismatch as $\Delta \ll \mu_{TI, S}$. Conservation of $k_{\parallel}$ (due to translational invariance) then gives for the angle of transmission in the superconductor: $\theta_{S}=\arcsin \left(\sin \theta \sqrt{\mu_{TI}^2-m_{z}^2}/ \mu_{S} \right)$.
Solving Eq. (\ref{eq1:zagoskin}) gives the energy as a function of the phase difference, $\phi$, between the superconductors. \\
\begin{figure*}[t!]
	\centering 
		\includegraphics[width=1.0\textwidth]{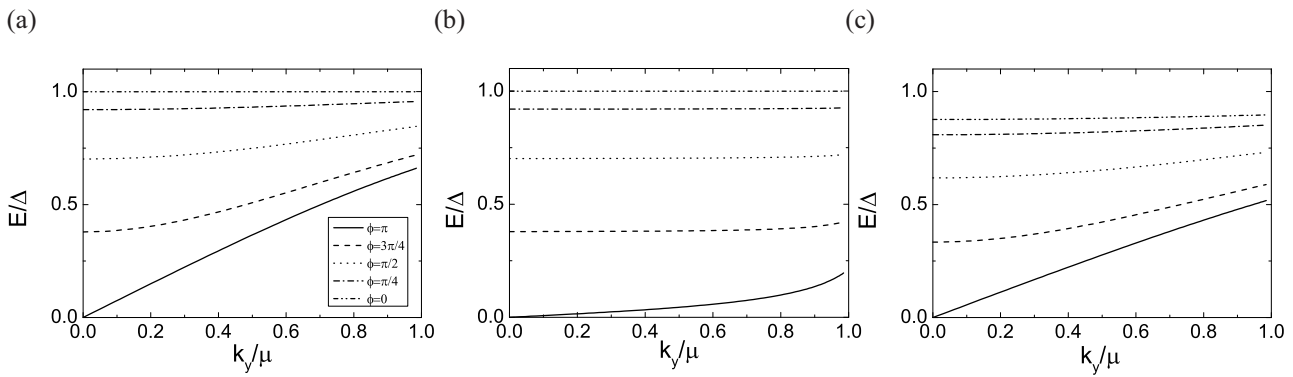}
		\caption{Bound state energy versus the wave vector parallel to the interface for different phase differences. For all the cases a nonhelical Majorana quantum wire is seen at $\phi=\pi$. The legend is shown in (a) and holds also for the other two. (a)$\mu_{TI}/\Delta=100, \mu_{S}/\Delta=1000$ and $L/L_{0}=0.01$. (b) $\mu_{TI}/\Delta=100, \mu_{S}/\Delta=110$ and $L/L_{0}=0.01$ and (c)$\mu_{TI}/\Delta=100, \mu_{S}/\Delta=1000, m_{z}/\Delta=60$ and $L/L_{0}=0.01$. } 
		\label{fig:3}
\end{figure*}

\section{Andreev bound states}
First, we consider the case with no ferromagnet on top of the TI. For perpendicular incidence we find a 4$\pi$ periodic ABS with a gapless dispersion, even in the presence of a momentum mismatch (the solid curves in Fig. \ref{fig:2}; the arrows in Fig. \ref{fig:2}(a) follow one ABS that is 4$\pi$ periodic). However, in the presence of a momentum mismatch, a nonzero angle of incidence results in a nonzero scattering amplitude and a gap is always present (Fig. \ref{fig:2} (a-c)). The larger the mismatch between the wave vectors, the larger is the gap that opens. For $\mu_{TI}=\mu_{S}$ the interface is effectively fully transparent and all trajectories give a 4$\pi$ periodic ABS. This is a consequence of the model where the superconducting gap is neglected. The opening of the gap at finite angles is due to finite back scattering at non-zero angle of incidence. For a larger mismatch between the chemical potentials, the difference in angles of the particle in the TI and superconductor is larger. This causes also a larger mismatch in the spin direction which increases the barrier and hence results in more reflected electrons. Only at zero angle of incidence, back scattering is prohibited by the topological nature. So in experiments no 4$\pi$ periodicity of the ABSs can be obtained for all angles; it is a single channel effect as is also concluded by Fu and Kane\cite{Fu2008} for a system with a small $\mu_{TI,S}$. Note, that even when this 4$\pi$ periodic ABS is present, it will only be noticeable in AC measurements since interactions with the environment already cause the system to reside in the lower (2$\pi$) ABS branches \cite{Badiane2011a,Pikulin2011,Fu2009,Wieder2012}. 

For a different length (Fig. \ref{fig:2} (c)) the curve corresponding to $\theta=2\pi/5$ is now lower in energy than the bound states of $\theta=\pi/4$ and $\theta=\pi/3$ compared to the graphs of Fig. \ref{fig:2}(a) and (b). In Fig. \ref{fig:2} (e) the bound state energy for specific angles at $\phi=\pi$ is plotted as function of length. We see an oscillating behavior as function of length due to a Fabry-Perot resonance. The oscillation period is determined by the Fabry-Perot resonant condition: $2Lk_{TI}\cos \theta=2\pi n$ with $k_{TI}$ the wave vector in the topological insulator and $n$ an integer\cite{Furusaki1991}. There is a strong similarity to a normal SNS junction. In Fig. \ref{fig:2} (f) a plot is made of the bound state energy of a SNS junction for a fixed angle of $\pi/4$ and phase difference of $\pi$ by determining the pole in the spectral supercurrent in Eq. (5) of Ref. \onlinecite{Furusaki1991} i.e.
\begin{eqnarray} 
\Gamma _{n}&= 0 = &\left(K^{2}\Omega^{2}_{n}+\omega^{2}_{n}\right)\cosh \left(\dfrac{2\omega_{n}L}{\hbar v_{n}}\right)+\nonumber \\
&& 2K \omega_{n}\Omega_{n}\sinh \left(\dfrac{2\omega_{n}L}{\hbar v_{n}}\right)-\nonumber \\
&&\left(K^{2}-1\right)\Omega^{2}_{n}\cos \left(2k_{N}L\right) +\Delta^{2}\cos \phi,
\end{eqnarray}where $\Omega_{n}=\sqrt{\omega_{n}^{2}+\Delta^{2}}$, $\omega_{n}=\pi\left(2n+1\right)/\beta$,  $\beta=1/k_{B}T$, $k_{N}=\sqrt{\frac{2m}{\hbar}\left(\mu-U\right)-\textbf{k}_{||}^{2}}$, $k_{S}=\sqrt{\frac{2m}{\hbar}\mu-\textbf{k}_{||}^{2}}$, $K=\frac{k^{2}_{N}+k^{2}_{S}}{2k_{N}k_{S}}$, $v_{N(S)}=\frac{\hbar k_{N(S)}}{m}$. $\Gamma_{n}=0$ corresponds with a pole in imaginary space which is equal to the energy of the Andreev bound state. The decrease of the amplitude is determined by the ratio of the length of the junction and the coherence length of the superconductor in the topological insulator. It should however be noted that varying the junction length will not result in a closing of the gap at a certain length. The calculation above is valid for every particular angle. When all angles are included the oscillations will be averaged out. \\
\\
When a ferromagnet is included, the magnetization is found to decrease the gap (see Fig. \ref{fig:2} (d)). This can be understood by considering the extreme case where the magnetization $m$ is close to the chemical potential so that the wave vectors of the electrons and holes are nearly zero (Eq. (6)). In that case it also follows from the conservation of $k_{\parallel}$ that $\theta_{s}$ is practically zero. Then by using the eigenvectors (Eq. (10), (11) and (17)) in the appendix and substitute these values of the wave vectors and angle into them, the resulting equations at the interfaces simplify. From these equations it can be seen that there is perfect Andreev reflection. Quantitatively it can be understood by noticing that the mismatch between the spins for the different particles in the system also causes a barrier. By the magnetization this mismatch becomes smaller due to the alignment.
 
In Fig. \ref{fig:3} (a) and (b) we show the bound states as a function of $k_{y}$, the wave vector parallel to the interface. For a phase difference of $\pi$ between the superconductors we see that the zero energy mode has a dispersion as function of $k_{y}$. This is also called a nonchiral Majorana state \cite{Fu2008}. For large mismatch between $k_{S}$ and $k_{TI}$ the model of Fu and Kane applies \cite{Fu2008} where the chemical potential is smaller than the superconducting gap. This situation resembles the case of a large mismatch at the interface in our case as the waves in the superconductor are then fully evanescent. For smaller mismatches the gap is smaller, which results in smaller slopes in the $k_{y}-E$ graphs. A smaller slope indicates a smaller velocity of this propagating mode along the interface as already noted in Ref. \onlinecite{Lababidi}. From the same reasoning as in the previous paragraph, the $k_{y}-E$ graphs have a smaller slope if magnetization is included. The result for $m_{z}/\Delta=60.0$ is shown in Fig. \ref{fig:3} (c).  

\begin{figure*}[t]
	\centering 
		\includegraphics[width=0.75\textwidth]{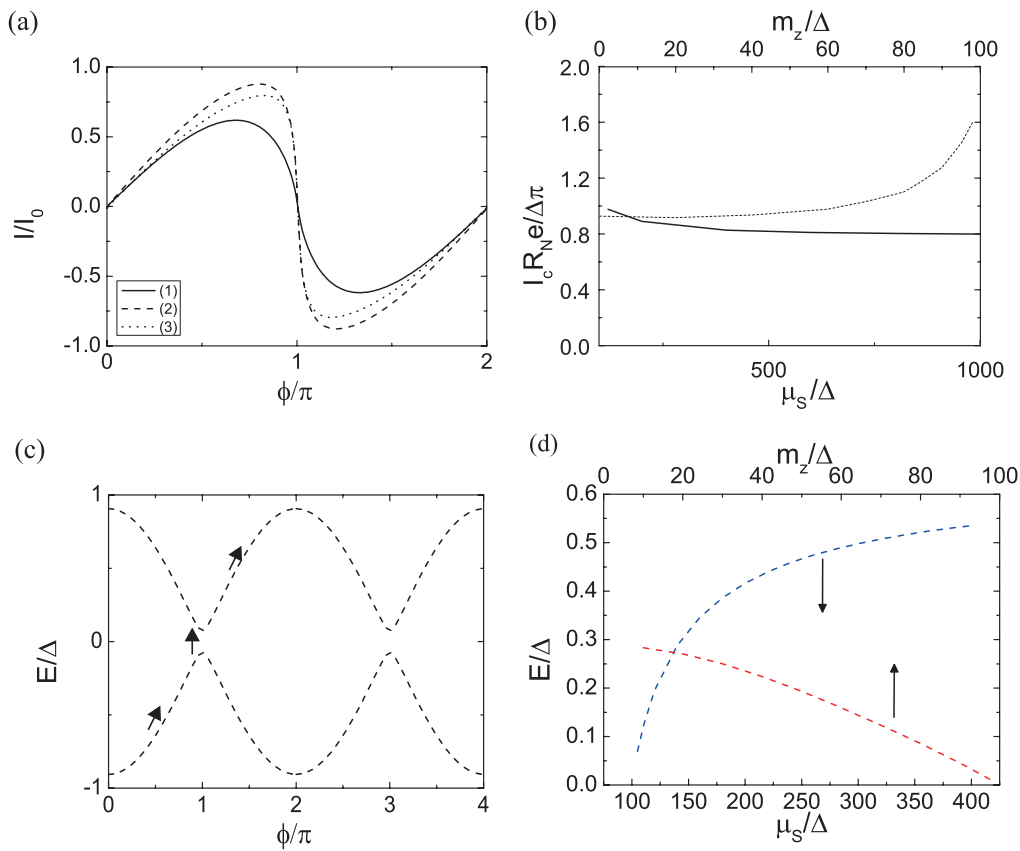}
		\caption{(a) Normalized Josephson supercurrent as a function of the phase difference between the superconductors. The numbers indicated in the inset correspond with the following parameters: (1) $\mu_{s}/\Delta=1000, \mu_{TI}/\Delta=100, m_{z}/\Delta=0, L/L_{0}=0.01$ ,(2)$\mu_{s}/\Delta=120, \mu_{TI}/\Delta=100, m_{z}/\Delta=0, L/L_{0}=0.01$ and (3) $\mu_{s}/\Delta=120, \mu_{TI}/\Delta=100, m_{z}/\Delta=60.0, L/L_{0}=0.01$. (b) Dependence of the normalized $I_{c}R_{N}e/\Delta$ product on both the magnetization (dashed line) and the chemical potential of the superconductor (solid line) separately. $\mu_{TI}/\Delta$ is kept constant to 100. In the dependence of the magnetization, $\mu_{S}/\Delta$ is kept constant to 120. The temperature is $T/T_{c}=0.01$ in (a) and (b). (c) Sketch of an Andreev bound state at non-zero angle of incidence. Due to magnetization the gap of the ABS has become so small that through Zener tunneling the electron in the lower branch can be promoted to the upper branch around $\phi=\pi \pm n2\pi$ where $n$ is an integer. (d) The influence of the magnetization and $\mu_{S}/\Delta$ on the value of the gap. For both situations $\mu_{TI}/\Delta=100, \theta=\pi/3$ and $L/L_{0}=0.01$.}
		\label{fig:4}
\end{figure*}
\section{Supercurrent}
In this section, we numerically calculate the angle-averaged supercurrent of the Andreev bound states. We consider here only the supercurrents for small junction length as longer lengths give no additional features regarding the 4$\pi$ periodic ABS and the influence of the chemical potential and magnetization on this. In the work of Ref. \onlinecite{Olund} discretized bound states were used but the continuum was missing in the calculation for larger length scales. Because we only consider here small length scales, only the discretized spectrum has to be considered
\begin{eqnarray}
I/I_{0}&=&\int_{-\pi/2}^{\pi/2} \! d\theta \cos \theta \tanh\left(\dfrac{E}{2k_{B}T}\right)\dfrac{dE/\Delta}{d\phi},
\end{eqnarray} where $I_{0}=eN\Delta/\hbar$.
Three plots of this normalized Josephson supercurrent are shown in Fig. \ref{fig:4} (a). The temperature is $T/T_{c}=0.01$ with $T_{c}$ the critical temperature. 

Although a 4$\pi$ periodicity is present in the Andreev bound state for zero angle of incidence, the other channels are 2$\pi$ periodic. Hence, the 2$\pi$ periodic character is dominating the angle-averaged supercurrent and therefore this current will be 2$\pi$ periodic in measurements. Moreover, the thermal equilibrium of the system even makes the ABS for zero angle of incidence 2$\pi$ periodic since inelastic scattering can relax quasiparticles to an ABS that is lower in energy\cite{Averin1996a,Averin1996b,Fu2009a}. This thermal equilibrium is due to the exchange between the bulk superconducting electrodes and the Andreev bound state levels in the junction\cite{Averin1996b}.

We see that for larger mismatch the supercurrent as function of phase has a more sinusoidal shape. For small mismatches there is a sharp transition at $\phi=\pi$ from positive to negative supercurrent which is also the case in a normal superconducting junction\cite{Golubov2004}. Also for a larger mismatch or for a magnetization the slope of the energy-phase curves become less steep, causing a smaller supercurrent for both cases. In Fig. \ref{fig:4} (b) we plot the dependence of the normalized $I_{c}R_{N}$ product as a function of mismatch in Fermi level and as function of magnetization. $R_{N}$ is here averaged over all angles. For larger mismatch in Fermi level the value of the $I_{c}R_{N}e/\Delta$ is saturating towards 0.5$\pi$ as expected in the tunneling limit. For small mismatch the value is $\pi$ as is also the case in the ballistic limit in normal SNS junctions. The critical current is also decreasing for a larger magnetization. However, we see that for values larger than $m_{z}/\Delta=60$ the critical current is increasing again. When analyzing the Andreev bound states we notice that there is a competition between the flatting of the bound states and the lowering of the barrier both due to magnetization. The latter depends on the relative magnitude of the magnetization to the mismatch. In Fig. \ref{fig:4} (b) the mismatch of the wave vectors due to a difference in the chemical potentials, is relatively small: $\mu_{TI}/\Delta=100$ and $\mu_{S}/\Delta=120$. The aligning of the spins for larger magnetization can therefore make the interface almost transparent. The corresponding Andreev bound states also resemble therefore an almost transparent interface: small gap and at $\phi=0$ and $2\pi$ the energy is $E/\Delta=\pm 1$. For a larger difference between $\mu_{S}$ and $\mu_{TI}$, the effect is less and the critical current is monotonically decreasing for larger magnetization. However, if we analyze the normal resistance, the resistance increases for larger magnetization. This is because the spins in the topological insulator with the ferromagnet on top are now more \textit{misaligned} compared to the topological insulator side without a ferromagnet on top. The combination of both an increasing $I_{c}$ and $R_{N}$, results in a normalized $I_{c}R_{N}e/\Delta$ product of $2\pi$. 

\section{Signatures of 4$\pi$ periodic ABS by means of Zener tunneling}
We have seen that the 4$\pi$ periodic ABS is a single channel effect for perpendicular trajectories only. Next to it, measuring in thermal equilibrium makes even the single 4$\pi$ ABS 2$\pi$ periodic because the electrons will follow the lowest ABS branches, i.e. below $E/\Delta=0$. The latter can be solved by doing AC measurements such as Shapiro step and/or noise measurements. The 4$\pi$ periodic ABS will only contribute to the Shapiro steps at a voltage equal to $n \hbar \omega/e$, where $n$ is an integer and $\omega$ the frequency of the applied microwave \cite{Badiane2011a,Fu2009}. A 2$\pi$ periodic ABS will result in Shapiro steps at $V=n \hbar \omega/2e$ which is half the step size of the 4$\pi$ periodic bound state.

However, due to the presence of just one single 4$\pi$ periodic ABS in 3D TIs/superconductor Josephson junctions out of many, one would expect that a 4$\pi$ periodic signature in AC measurements is not visible due to angle-averaging. Usually, one can enhance the zero angle contribution by introducing a physical barrier of finite width due to the exponential dependence of the wave function on the width. It is however not possible to cancel the nonzero angles by introducing a physical barrier between the superconductor and TI because of Klein tunneling which renders barriers effectively transparent (see for example the discussion of Klein tunneling in graphene in Ref. \onlinecite{Katsnelson2006,BeenakkerGraphene}). So in order to see a 4$\pi$ periodic ABS, the ABSs of all angles should be 4$\pi$ periodic. By means of Zener tunneling this can be achieved. 

In order to obtain Zener tunneling, a bias voltage across the junction is required (which is for example the case in Shapiro steps measurements). Due to this bias voltage the quasi-particles in the junction can gain enough kinetic energy so that they can transfer from a lower Andreev bound state to an upper bound state despite the separation by a gap\cite{Kroemer1999,Jacobs2005}. It is noted in Ref. \onlinecite{Averin1995} that for large transparency of the interface this can result in a $4\pi$ periodic ABS. So when the gap is small, it cannot be distinguished anymore from an Andreev bound state without a gap. Hence, when electrons in the lower branch of these Andreev bound states with small gaps are promoted to the upper branch, it can result in $4\pi$ periodic signatures in AC measurements \cite{Sau2012,Pikulin2012}. In Fig. \ref{fig:4}(c) a sketch is shown of an ABS with a finite but small gap. Due to magnetization this can result in a 4$\pi$ periodic signature in for example Shapiro step and/or noise measurements. Physically the small gap is similar to having a finite length in the 1D Kitaev model where also a gap is present due to the interaction of the Majorana fermions at the ends \cite{Pikulin2011,Kitaev,Pikulin2011b,Jose2012}. A way to reduce the gaps of all non-zero angle of incidence channels in a 3D TI in order to enhance the chance of Zener tunneling to get the $4\pi$ periodicity of all ABSs, is by exploiting magnetization as is shown in Fig. \ref{fig:2}(d).

The influence of the magnitude of the magnetization depends first of all on the relative magnitude of the chemical potentials to each other as we can see from Fig. \ref{fig:4}(d). The larger the mismatch the less the influence is of the magnetization. Secondly, the influence of the magnetization depends on its magnitude compared to the absolute magnitude of the chemical potential. With a larger chemical potential, the influence of the magnetization is less. To get a clearer picture of the influence of the magnetization, we have plotted the gap in the Andreev bound state (at $\phi=\pi$) for several conditions in Fig. \ref{fig:4} (d). For an increase of mismatch we kept the magnetization constant to $m_{z}/\Delta=0$, length $L/L_{0}=0.01$ and angle $\theta=\pi/3$. A similar result is obtained for other angles. Further $\mu_{TI}/\Delta=100$ for both graphs. We see that the gap is a strong function of magnetization in the beginning but saturates at larger mismatches in the chemical potentials. The graph that shows the influence of the magnetization has a constant (small) mismatch $\mu_{S}/\Delta=140$ and $\mu_{TI}/\Delta=100$. When the magnetization energy is half the value of the Fermi level in the TI, the energy gap is already decreased by 50\% of its original value at zero magnetization. However we have estimated before that the magnetization energy is typically on the order of 1\% of the Fermi energy, i.e. $\mu \gg m_{z}$ in experiments. So only for small mismatches in the chemical potentials it is possible to reduce the gap so that by means of Zener tunneling the 4$\pi$ periodicity of the ABSs remains for all angles. This reduction of the gap can be made visible in SQUID experiments as proposed in Ref. \onlinecite{Veldhorst2012c,Veldhorst2012d}, where it is shown that the reduction of the gap size results in a different critical current modulation of the SQUID as function of the applied flux through the ring. These results hold even in equilibrium experiments, so that these SQUID devices can be used for ABS spectroscopy. 

\section{Conclusions}
We studied the Andreev bound states and the resulting supercurrent for S/TI/S junctions with and without a ferromagnet on top of the TI. In experiments it is often the case that $\mu\gg \Delta$ and $m_{z}<\mu$ and therefore we extended the model (by Fu and Kane \cite{Fu2008} and Linder \textit{et al.}\cite{Linder2010}) towards this regime. The bound states are solved by means of the total wave function in the topological insulator and its relation to the reflection coefficients, providing insight in the process. The important conclusion is that the results from Fu and Kane \cite{Fu2008} (e.g. the 4$\pi$ periodic ABS existing only for $\theta=0$) still remain, even for large chemical potentials. Therefore, these features are also valid in actual experiments on TIs. The 4$\pi$ periodicity of the bound states only remains in a 3D topological insulator for zero angle of incidence. This 4$\pi$ feature cannot be observed as all the other angles, which give 2$\pi$ periodic bound states, causes the 2$\pi$ periodicity to dominate. However, Zener tunneling can cause a transition from the lower branches to the upper branches of the Andreev bound states, even when a gap is present. We can enhance this Zener tunneling, and hence enhancing the 4$\pi$ periodicity of the non-zero angle ABSs, by depositing a ferromagnet on top. The magnetization effectively lowers the barrier which causes the gap in the Andreev bound states to become smaller. 

This work is supported by the Netherlands Organization for Scientific Research (NWO) by the Dutch Foundation for Fundamental Research on Matter (FOM). 

\appendix
\section*{Appendix: Eigenvectors and boundary conditions}
The eigenvectors in the topological insulator are calculated to be
\begin{eqnarray}
\psi_{1}&=&n_{1}\left(\begin{array}{c}
-m+\sqrt{m^{2}+v^{2}|\textbf{k}_{1}|^{2}}\\
-v|\textbf{k}_{1}|e^{i\theta}\\
0\\
0
\end{array} \right) ,\\
\psi_{2}&=&p_{2}\left(\begin{array}{c}
0\\
0\\
m+\sqrt{m^{2}+v^{2}|\textbf{k}|_{2}^{2}}\\
-v|\textbf{k}_{2}|e^{-i\theta}
\end{array} \right) ,\\
\psi_{3}&=&p_{3}\left(\begin{array}{c}
m+\sqrt{m^{2}+v^{2}|\textbf{k}_{3}|^{2}}\\
v|\textbf{k}_{3}|e^{i\theta}\\
0\\
0
\end{array} \right) ,\\
\psi_{4}&=&n_{4}\left(\begin{array}{c}
0\\
0\\
-m+\sqrt{m^{2}+v^{2}|\textbf{k}_{4}|^{2}}\\
v|\textbf{k}_{4}|e^{-i\theta}
\end{array} \right),
\end{eqnarray} where $n_{j}=1/\sqrt{2(m^{2}+v^{2}|\textbf{k}_{j}|^{2}-m\sqrt{m^{2}+v^{2}|\textbf{k}_{j}|^{2}})}$ with $j=1,4$ for $\psi_{1}$ and $\psi_{4}$ respectively. Further $p_{j}=1/\sqrt{2(m^{2}+v^{2}|\textbf{k}_{j}|^{2}+m\sqrt{m^{2}+v^{2}|\textbf{k}_{j}|^{2}})}$ with $j=2,3$ for $\psi_{2}$ and $\psi_{3}$ respectively. The eigenvalues are
\begin{eqnarray}
E_{1}&=&-\mu-\sqrt{m^{2}+v^{2}|\textbf{k}_{1}|^{2}},\\
E_{2}&=&\mu-\sqrt{m^{2}+v^{2}|\textbf{k}_{2}|^{2}},\\
E_{3}&=&-\mu+\sqrt{m^{2}+v^{2}|\textbf{k}_{3}|^{2}},\\
E_{4}&=&\mu+\sqrt{m^{2}+v^{2}|\textbf{k}_{4}|^{2}}.
\end{eqnarray} $\psi_{1}$ and $\psi_{3}$ are the electrons belonging to the lower and upper half of the Dirac cone respectively. $\psi_{2}$ and $\psi_{4}$ are the holes corresponding respectively to the lower and upper half of cone. The wavefunction in the superconductor is given by
\begin{eqnarray}
\psi_{s}&=&\dfrac{1}{2\sqrt{E}}\left( 
\begin{array}{c}
e^{i\phi}\sqrt{E-\mu+v|\textbf{k}_{s}|}\\
e^{i\phi}e^{i\theta}\sqrt{E-\mu+v|\textbf{k}_{s}|}\\
\dfrac{-\Delta e^{i\theta}}{\sqrt{E-\mu+v|\textbf{k}_{s}|}}\\
\dfrac{\Delta}{\sqrt{E-\mu+v|\textbf{k}_{s}|}}
\end{array}\right) 
\end{eqnarray} with an energy given by $E=\sqrt{\Delta^{2}+(v|\textbf{k}_{s}|- \mu)^{2}}$. 

The reflection and transmission coefficients can differ at both interfaces due to magnetization. If we consider the electrons at the upper cone as depicted in Fig. \ref{fig:1} (a) we need the wave functions $\psi_{2}$ and $\psi_{3}$. Taking the direction of the particles into account in the angle $\theta$ in the TI and $\theta_{s}$ in the superconductor we arrive at the following set of equations:\\
\textbf{Right interface, incoming electron}
\begin{eqnarray}
p_{m,3}\left(1+r_{ee}\right) &=& \dfrac{e^{i\phi}}{2\sqrt{E}}\left(c_{1}t_{ee}+d_{1}t_{eh} \right) ,\nonumber \\
p_{k,3}\left(e^{i\theta}-r_{ee}e^{-i\theta}\right) &=& \dfrac{e^{i\phi}}{2\sqrt{E}}\left(c_{1}t_{ee}e^{i\theta_{s}}-d_{1}t_{eh}e^{-i\theta_{s}} \right),\nonumber\\
p_{m,2}r_{eh}&=&\dfrac{\Delta}{2\sqrt{E}}\left(-t_{ee}\dfrac{e^{i\theta_{s}}}{c_{2}}+t_{eh}\dfrac{e^{-i\theta_{s}}}{d_{2}} \right),\nonumber\\
-p_{k,2}e^{-i\theta}r_{eh}&=&\dfrac{\Delta}{2\sqrt{E}}\left(t_{ee}\dfrac{1}{c_{2}}+t_{eh}\dfrac{1}{d_{2}} \right),
\end{eqnarray} where 
\begin{eqnarray}
|\textbf{k}_{se}|&=&\mu/v+\sqrt{E^{2}-\Delta^{2}}/v, \nonumber \\
|\textbf{k}_{sh}|&=&\mu/v-\sqrt{E^{2}-\Delta^{2}}/v,\nonumber \\
c_{1}&=&\sqrt{E-\mu+v|\textbf{k}_{se}|},\nonumber \\
d_{1}&=&\sqrt{E-\mu+v|\textbf{k}_{sh}|},\nonumber \\
c_{2}&=&\sqrt{E-\mu+v|\textbf{k}_{se}|},\nonumber \\
d_{2}&=&\sqrt{E-\mu+v|\textbf{k}_{sh}|},\nonumber \\
p_{m,j}&=&\dfrac{m+\sqrt{m^{2}+v^{2}|\textbf{k}_{j}|^{2}}}{{\sqrt{2(m^{2}+v^{2}|\textbf{k}_{j}|^{2}+m\sqrt{m^{2}+v^{2}|\textbf{k}_{j}|^{2}})}}},\nonumber \\
p_{k,j}&=&\dfrac{v|\textbf{k}_{j}|}{\sqrt{2(m^{2}+v^{2}|\textbf{k}_{j}|^{2}+m\sqrt{m^{2}+v^{2}|\textbf{k}_{j}|^{2}})}}.
\end{eqnarray}\\
\textbf{Right interface and incoming hole }
\begin{eqnarray}
p_{m,3}r_{he}&=&\dfrac{e^{i\phi}}{2\sqrt{E}}\left(c_{1}t_{ee}+d_{1}t_{eh}\right),\nonumber\\
-p_{k,3}r_{he}e^{-i\theta}&=&\dfrac{e^{i\phi}}{2\sqrt{E}}\left(c_{1}t_{ee}e^{i\theta_{s}}-d_{1}t_{eh}e^{-i\theta_{s}}\right),\nonumber\\
p_{m,2}\left(1+r_{hh}\right)&=&\dfrac{\Delta}{2\sqrt{E}}\left(-t_{ee}\dfrac{e^{i\theta_{s}}}{c_{2}}+t_{eh}\dfrac{e^{-i\theta_{s}}}{d_{2}} \right),\nonumber\\
p_{k,2}\left(e^{i\theta}-r_{hh}e^{-i\theta}\right)&=&\dfrac{\Delta}{2\sqrt{E}}\left(t_{ee}\dfrac{1}{c_{2}}+t_{eh}\dfrac{1}{d_{2}} \right).
\end{eqnarray}\\
\textbf{Left interface, incoming electron}
\begin{eqnarray}
p_{m,3}\left(1+r_{ee} \right)&=&\dfrac{e^{i\phi}}{2\sqrt{E}}\left(d_{1}t_{eh}+c_{1}t_{ee} \right),\nonumber\\
p_{k,3}\left(-e^{-i\theta}+r_{ee}e^{i\theta} \right)&=&\dfrac{e^{i\phi}}{2\sqrt{E}}\left(d_{1}t_{eh}e^{i\theta_{s}}-c_{1}t_{ee}e^{-i\theta_{s}}\right),\nonumber\\
p_{m,2}r_{eh}&=&\dfrac{\Delta}{2\sqrt{E}}\left(-t_{eh}\dfrac{e^{i\theta_{s}}}{d_{2}}+t_{ee}\dfrac{e^{-i\theta_{s}}}{c_{2}} \right),\nonumber\\
p_{k,2}r_{eh}e^{i\theta}&=&\dfrac{\Delta}{2\sqrt{E}}\left(t_{eh}\dfrac{1}{d_{2}}+t_{ee}\dfrac{1}{c_{2}} \right).
\end{eqnarray}\\
\textbf{Left interface, incoming hole}
\begin{eqnarray}
p_{m,3}r_{he}&=&\dfrac{e^{i\phi}}{2\sqrt{E}}\left(d_{1}t_{eh}+c_{1}t_{ee}\right),\nonumber\\
p_{k,3}r_{he}e^{i\theta}&=&\dfrac{e^{i\phi}}{2\sqrt{E}}\left(d_{1}t_{eh}e^{i\theta_{s}}-c_{1}t_{ee}e^{-i\theta_{s}}\right),\nonumber\\
p_{m,2}\left(1+r_{hh} \right)&=&\dfrac{\Delta}{2\sqrt{E}}\left(-t_{eh}\dfrac{e^{i\theta_{s}}}{d_{2}}+t_{ee}\dfrac{e^{-i\theta_{s}}}{c_{2}} \right),\nonumber\\
p_{k,2}\left(-e^{-i\theta}+r_{hh}e^{i\theta} \right)&=&\dfrac{\Delta}{2\sqrt{E}}\left(t_{eh}\dfrac{1}{d_{2}}+t_{ee}\dfrac{1}{c_{1}} \right) .
\end{eqnarray}
These equations can be used to calculate the coefficients that are used in Eq. (\ref{eq1:zagoskin}) in the main part of the paper.

\end{document}